\documentclass{article}

\usepackage{PRIMEarxiv}

\usepackage{amsmath,amsfonts,bm}

\def\eqref#1{equation~\ref{#1}}

\def\1{\bm{1}}

\DeclareMathAlphabet{\mathsfit}{\encodingdefault}{\sfdefault}{m}{sl}
\SetMathAlphabet{\mathsfit}{bold}{\encodingdefault}{\sfdefault}{bx}{n}

\usepackage[utf8]{inputenc} 
\usepackage[T1]{fontenc}    
\usepackage{hyperref}       
\usepackage{url}            
\usepackage{booktabs}       
\usepackage{amsfonts}       
\usepackage{nicefrac}       
\usepackage{microtype}      
\usepackage{lipsum}
\usepackage{fancyhdr}       
\usepackage{graphicx}       
\graphicspath{{media/}}     

\usepackage{cleveref}
\usepackage{enumitem}
\usepackage{wrapfig}
\usepackage{subfigure}

\usepackage[numbers]{natbib}

\usepackage{CJKutf8}
\usepackage{cleveref}
\usepackage{graphicx}
\usepackage{booktabs}
\usepackage{subfigure}
\usepackage{multirow}
\usepackage{makecell}
\usepackage{enumitem}
\usepackage{xcolor, color, soul}

\definecolor{lightblue}{RGB}{219,226,238}
\definecolor{darkred}{RGB}{105,17,10}
\definecolor{lightyellow}{RGB}{251,242,214}
\definecolor{darkyellow}{RGB}{82,58,34}
\definecolor{lightgrey}{RGB}{230,230,230}
\definecolor{darkgrey}{RGB}{57,57,57}

\newcommand{\hlblue}[1]{\sethlcolor{lightblue}\textit{\textbf{\color{blue}{\hl{#1}}}}}

\newcommand{\hlyellow}[1]{\sethlcolor{lightyellow}\textit{\textbf{\color{darkyellow}{\hl{#1}}}}}

\title{Music-to-Text Synaesthesia: Generating Descriptive Text from Music Recordings}

\author{
  Zhihuan Kuang$^1$, Shi Zong$^1$, Jianbing Zhang$^1$, Jiajun Chen$^1$, Hongfu Liu$^2$ \\
  $^1$Nanjing University \quad   $^2$Brandeis University \\
  \texttt{kuangzh@smail.nju.edu.cn} \quad
  \texttt{\{szong, zjb, chenjj\}@nju.edu.cn} \\  
  \texttt{hongfuliu@brandeis.edu}
}

\begin{document}

\maketitle

\vspace{-.2cm}

\begin{abstract}
In this paper, we consider a novel research problem: music-to-text synaesthesia. Different from the classical music tagging problem that classifies a music recording into pre-defined categories, music-to-text synaesthesia aims to generate descriptive texts from music recordings with the same sentiment for further understanding.  
As existing music-related datasets do not contain the semantic descriptions on music recordings, we collect a new dataset that contains 1,955 aligned pairs of classical music recordings and text descriptions. 
Based on this, we build a computational model to generate sentences that can describe the content of the music recording. 
To tackle the highly non-discriminative classical music, we design a group topology-preservation loss, which considers more samples as a group reference and preserves the relative topology among different samples. 
Extensive experimental results qualitatively and quantitatively demonstrate the effectiveness of our proposed model over five heuristics or pre-trained competitive methods and their variants on our collected dataset.\footnote{Our code are available at \url{https://github.com/MusicTextSynaesthesia}.}
\end{abstract}

\section{Introduction}
\label{sec:intro}

Our physical world is naturally composed of various modalities.
In recent years, multi-modal learning has drawn great attention and has been developed in diverse applications.
Visual frames in videos are matched with text captions and these pairs have been widely used for video-language pre-training \citep{Sun_2019_ICCV, 10.1145/3474085.3475703, li-etal-2020-hero}; Kinects employ the RGB camera and the depth sensor for action recognition and human pose estimation \citep{10.1109/CVPR.2011.5995316, Carreira_2017_CVPR}; autonomous driving cars integrate the visible and invisible lights by the camera, radar, and lidar for a series of driving-related tasks \citep{10.5555/1554715, 8852086}; cross-modal retrieval aims to match text with the existing textual repository and other modalities to meet users' queries \citep{Nagrani18a, Suris_2018_ECCV_Workshops, Zeng_2021_CVPR}; language grounding learns the meaning of language by leveraging the sensory data such as video or images \citep{bisk-etal-2020-experience, thomason2021language}.

Besides the above studies that employ multi-modal data to jointly achieve the learning task, translating information among different modalities, also regarded as synaesthesia, is another crucial task in the multi-modal community.
Various methods for synaesthesia between text and other modalities have been studied. 
Speech recognition can be directly regarded as a translation between the text and audio modality \citep{shen_natural_2018}. 
Image captioning extracts the high-level visual cues and translates them into a descriptive sentence to describe the image content, while some studies consider the inverse process of image captioning by converting a semantic text into the visual image \citep{huang_unifying_2021, xu_show_2015}. Different from the existing modality translation studies, in this paper, we consider a novel problem, \textit{music-to-text synaesthesia}, i.e., generating descriptive texts from music recordings with the same sentiment orientation.

\begin{figure*}[t]
    \centering
    \includegraphics[width=.99\textwidth]{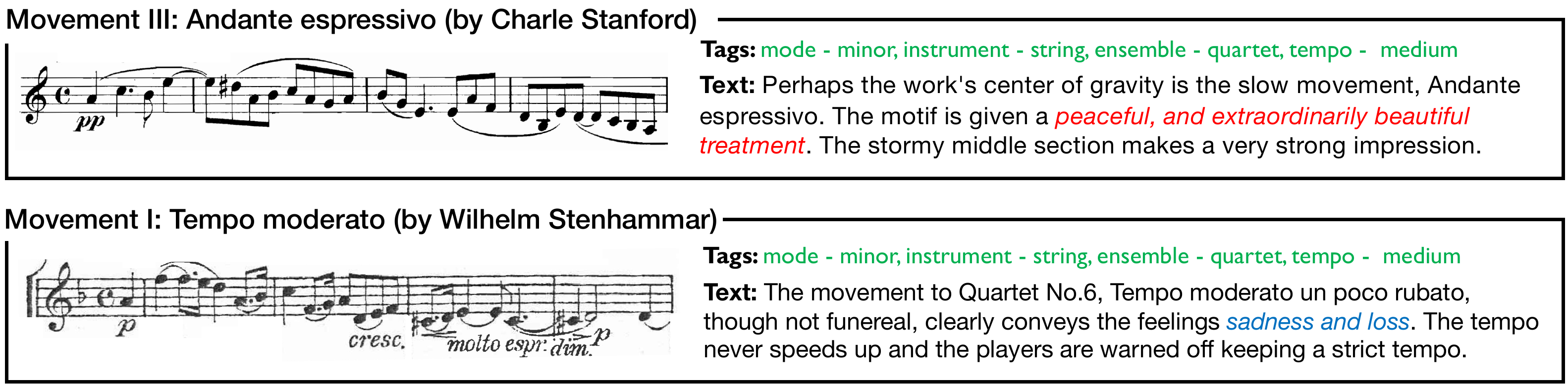}
    \caption{Samples of classical music and corresponding  descriptions in our collected dataset. The first piece is from String Quartet No.2 in A minor Op.45 composed by Charles Stanford, and the second is from String Quartet No.6 in D minor Op.35 by Wilhelm Stenhammar.
    These two samples have the same music tags but different sentiments.
    }
    \label{fig:example}
\end{figure*}

There have been some pioneering attempts that build the connections between music recordings and tags at the initial stage.
\citet{cai-etal-2020-music} formulates music auto-tagging as a captioning task and automatically outputs a sequence of tags given a clip of music.
\citet{zhang-etal-2020-butter} uses keywords of music key, meter and style to generate music descriptions, which can be used for caption generation.
However, we argue that descriptive texts contain much richer information than tags, thus providing a better understanding of music recording. 
Moreover, we notice that tags might have a biased interpretation. 
To demonstrate this, in \Cref{fig:example} we present two music recordings with the same music tags, but the opposite sentiment orientation of the text.  
The first one expresses a positive sentiment by describing the music as ``\textit{peaceful}'' and ``\textit{beautiful},'' while the second one uses tokens including ``\textit{sadness}'' and ``\textit{loss}'' to express a negative sentiment. It is clear that music tags are insufficient for describing the content of a music piece.

\textbf{Contributions}. In this paper, we propose a new task of generating descriptive text from music recordings. Specifically, given a music recording, we aim to build a computational model that can generate sentences that can describe the content of the music recording, as well as the music's inherent sentiment. 
We make the following contributions:

\begin{itemize}[wide=10pt, leftmargin=*]
    \item From the research problem perspective, different from the music tagging problem, our proposed music-to-text synaesthesia is a cross-modality translation task that aims at converting a given music piece to a text description. 
    To our best knowledge, it is a novel research problem in the multi-modal learning community.
    \item From the dataset perspective, the existing music-related datasets do not contain the semantic description of music recordings. To build computational models for this task, we collect a new dataset that contains 1,955 aligned pairs of classical music recordings and text descriptions.
    \item From the technical perspective, we design a group topology-preservation loss in our computational model to tackle the non-discriminative music representation, which considers more data points as a group reference and preserves the relative topology among different nodes. Thus it can better align the music representations with the structure in text space.
    \item From the empirical evaluation, extensive experimental results demonstrate the effectiveness of our proposed model over five heuristics or pre-trained competitive methods and their variants on our collected dataset. We also provide several case studies for comparisons and elaborate the explorations on our group topology-preservation loss and some parameter analyses.
\end{itemize}

\section{Related Work}
\label{sec:related_work}
We introduce the related work on multi-modality learning and music tagging and captioning below.

\paragraph{Multi-modality Learning.}
The goal of multi-modal machine learning is to build computational models that are able to process and relate information from different modalities, such as audio, text, and image. 
A large portion of prior works has focused on modality fusion, which aims at making predictions by joining information from two or more modalities \cite{10.1109/TPAMI.2018.2798607}. 
Applications include audio-visual speech recognition \citep{8585066}, visual question answering \citep{balanced_vqa_v2}, and media summarization.

Beyond multi-modality fusion, translation among different modalities also draws increasing attention. 
There are three common frameworks of multi-modality translation.
(1) Encoder-decoder models directly learn intermediate representations used for projecting one modality into another.
\citet{Zhang_2017_ICCV} adapts a sketch-refinement process to generate photo-realistic images for text-to-image synthesis tasks.
\citet{Wang_2021_ICCV} designs a framework for end-to-end dense video captioning with parallel decoding.
(2) Models with joint representations fuse multi-modal features by mapping representations of different modalities together into a shared semantic subspace. 
\citet{Sun_2019_ICCV} proposes ViLBERT, which extends BERT architecture to a multi-modal two-stream model, which learns task-agnostic joint representations of image content and natural language.
\citet{7740886} designs an embedding between video features and term vectors to learn the entire representation from freely available web videos and their descriptions.
(3) Representations in coordinated representations-based models exist in separated spaces, but are coordinated through a similarity function (e.g., Euclidean distance) or a structure constraint. 
These works include \citet{10.1145/3123266.3123326}, which present a method to learn a common subspace based on adversarial learning for adversarial cross-modal retrieval.
\citet{10.1109/TIP.2018.2852503} proposed a modality-specific cross-modal similarity measurement approach for tasks including cross-modal retrieval.
In this work, we experiment with different losses on the coordinate model, as it achieves the best performance among all three different types of models.

\paragraph{Music Tagging and Captioning Tasks.}
We notice some pioneering studies on music tagging or captioning tasks~\citep{Choi2016TowardsMC, 9859499, Manco2022ContrastiveAL}. 
\citet{9533461} uses a private production music dataset, with music clips of length between only 30 and 360 seconds and captions containing between 3 and 22 tokens.
Their proposed model is an encoder-decoder network with a multimodal CNN-LSTM encoder with temporal attention and an LSTM decoder.
Our proposed task is different from tagging and captioning tasks, as we aim at translating semantics and preserving sentiment between modalities. 

Existing public music-related datasets mainly contain simple music tags.
\textit{AudioSet} dataset \citep{45857} is a large-scale collection of human-labeled 10-second sound clips (not music recording) drawn from YouTube videos. 
This dataset only has descriptions for categories, not for individual sounds.
The \textit{MTG-Jamendo} dataset \citep{bogdanov2019mtg} contains over 55,000 full audio tracks with 195 tags ranging from genre, instrument, and mood/theme categories.
\citet{3490} describes a dataset containing reviewers from Amazon for albums. However, users' reviews may not necessarily describe the actual contents of music recordings.
\citet{cai-etal-2020-music} formulates the music tagging problem as a multi-class classification problem. A dataset called \textit{MajorMiner} is used, with each music recording associated with tags collected from different users.
\citet{zhang-etal-2020-butter} studies bidirectional music-sentence retrieval and generation tasks. The used dataset contains 16,257 folk songs paired with metadata information, including select key, meter, and style as keywords. 
However, text describing music only focuses on specific information and has limited writing styles.

\section{Data Collection and Analysis}
\label{sec:data_collection}
In order to generate descriptive texts from music recordings,  a dataset containing aligned music-text pairs is required for model training. Although there are several public music/audio datasets with tags or user reviews (see Section~\ref{sec:related_work}), unfortunately, they are not suitable for our task for the following reasons: (1) From the text side, current datasets only have pre-defined tags for music pieces, rather than descriptive texts for music contents. (2) From the audio side, some clips are too short without a musical melody. In light of this, we collect a new dataset for music-to-text synaesthesia task.

\paragraph{Data Collection and Post-Processing.}
We collected the data from EARSENSE,\footnote{\scriptsize{\url{http://earsense.org/}}} a website that hosts a database for chamber music.
EARSENSE provides comprehensive meta-information for each music composition, including composers, works, and related multi-media resources. There is also an associated introductory article from professional experts, with detailed explanations, comments or analyses for movements.  
Figure~\ref{fig:example} shows an illustrative example of the music-text pairs. A typical music composition contains several \textit{movements}. Each movement has its own title that normally contains tempo markings or terms such as minuet and trio; in some cases, it has a unique name speaking to the larger story of the entire work. As movements have their own form, key, and mood, and often contain a complete resolution or ending, we will treat each movement as the basic unit in this work.\footnote{For example, Ludwig van Beethoven's Sonata Pathétique (No. 8 in C minor, Op. 13) contains three movements: (I) Grave (slowly, with solemnity), (II) Adagio cantabile (slowly, in a singing style), and (III) Rondo: Allegro (quickly).}

\begin{wrapfigure}[12]{r}{3.2in}\vspace{-6mm}
    \centering
    \subfigure[music similarity]{
    \centering
    \includegraphics[width=.24\textwidth]{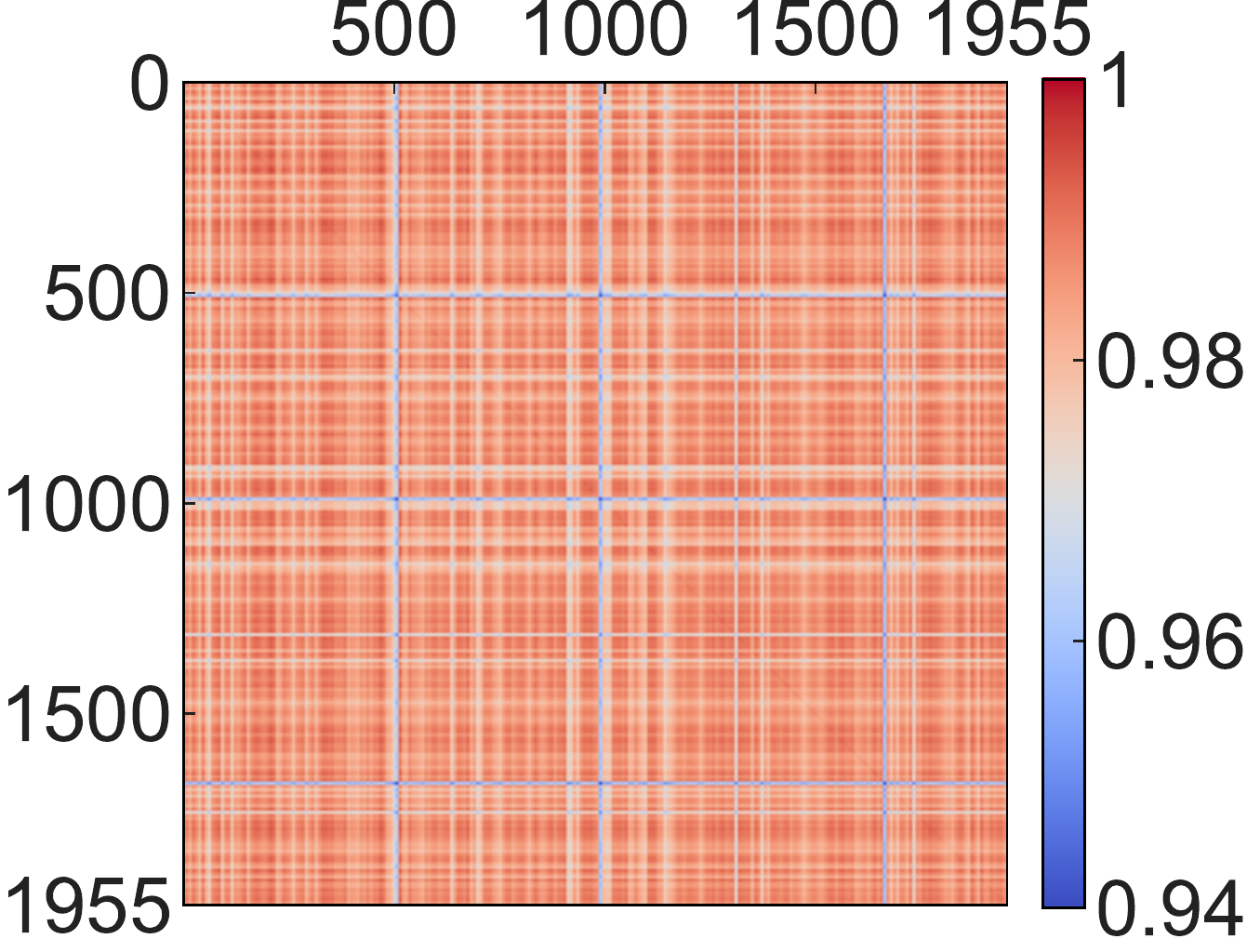}
    }%
    \subfigure[text similarity]{
    \centering
    \includegraphics[width=.24\textwidth]{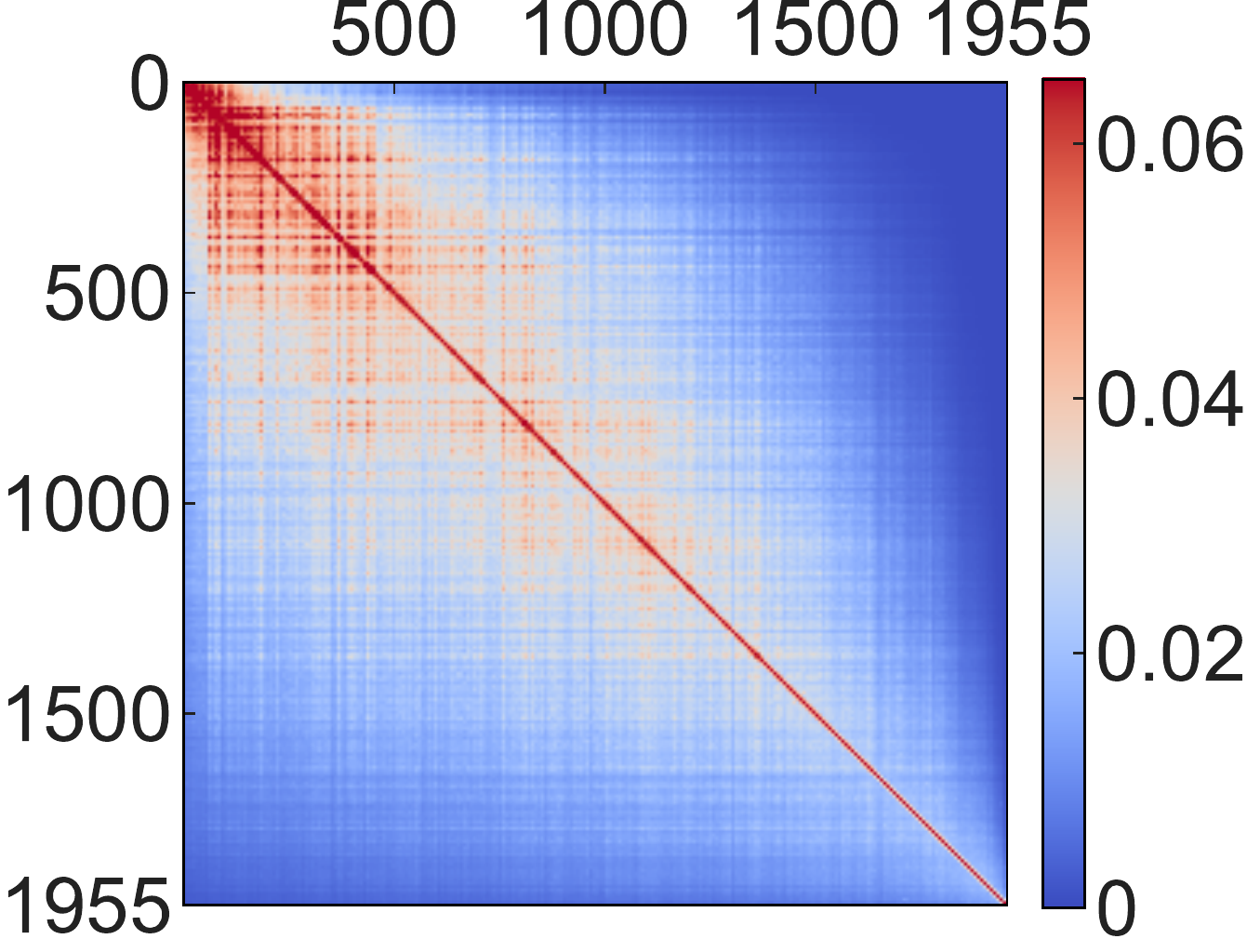}\vspace{-2mm}
    }\vspace{-2mm}
    \caption{Pairwise similarity matrices of music representation by a self-reconstruction autoencoder and raw text by cosine and BLEU score.}
    \label{fig:similarity}
\end{wrapfigure}

We managed to collect 2,380 text descriptions in total, where 1,955 descriptions have corresponding music pieces.
We converted the tempo markings in the titles of movements into universal four categories from slow to super fast.
These categories are then added for movements as tags, by directly checking whether it contains tokens in our list.

\paragraph{Preliminary Exploration.}
We observe that the lengths of the 95\% collected music pieces vary from 2.5 to 14 minutes, which correspond to the descriptive texts with 14 to 192 tokens.
We provide more details of our data statistics in \Cref{sec:data_stats}.

We further explore the pairwise similarity and check whether similar music pieces share similar textual descriptions. 
In Figure~\ref{fig:similarity}, we plot the pairwise similarity matrix of music and text,
where we use the cosine similarity on the music representations derived from a self-reconstruction auto-encoder and use BLEU score to calculate the raw text similarity. 
We notice that the texts are highly discriminative, where 96\% of the pairwise similarities are below 0.06. However, all music representations have over 0.95 similarity due to the high similarities among classical music. It inevitably increases the difficulty of learning a mapping function for the music-to-text synaesthesia task that similar music pieces match divergent texts. 

To tackle the above challenge, we aim to employ the text to guide music representations and make them more discriminative, so the cross-modality translation would be easier.  In the next section, we elaborate our proposed computational model for music-to-text synaesthesia.

\section{Model}\label{sec:model}

Our music-to-text synaesthesia is a cross-modality translation problem. Given $n$ paired tuple $\langle m_i, t_i\rangle$ with $1$$\le$$i$$\le$$n$, where $m_i$ denotes the tuple $i$'s music and $t_i$ for its corresponding text, a music-to-text synaesthesia model $\mathcal{M}\to\mathcal{T}$ builds a mapping from the music to the text space, where $\mathcal{M}$ and $\mathcal{T}$ present the music and text space, respectively. 
In this section, we first provide a general overview of our model structure, followed by our proposed group topology-preservation loss to tackle the non-discriminative music representation. 

\paragraph{Cross-Modality Translation Model.}
Our translation model (in \Cref{fig:framework}) is mainly based on the coordinate model~\citep{10.1109/TIP.2018.2852503}. Our proposed group topology-preservation loss is then used to achieve the translation between two modalities by learning the music/text latent space and their mapping by using three independent auto-encoders. 

\textit{Learning Music Representation}. We choose a convolutional neural network structure as our music feature extractor. Specifically, we use the spectrum of the raw music as inputs, where a CNN encoder and a transposed-CNN decoder are connected for the music reconstruction task. Let $f(\cdot)$ as the music encoder and $f'(\cdot)$ as the music decoder. We have the loss function for music representation learning as follows:
\begin{equation}\nonumber
\small
\ell_{m2m} = \sum\nolimits_{i=1}^n ||{m_i} - f'(f({m_i}))||^2.
\end{equation}

\textit{Learning Text Representation.}
Similarly, in the text field, a transformer-based encoder and decoder are combined to train a text reconstruction model. Let $g(\cdot)$ be the text encoder and $g'(\cdot)$ be the text decoder. 
We have the loss function for text representation learning as follows: 
\begin{equation}\nonumber
\small
\ell_{t2t} = \sum\nolimits_{i=1}^n \text{CrossEntropy}(t_i, g'(g(t_i))).
\end{equation}

\textit{Transforming From Music to Text.} 
With the above music and text representation, an extra auto-encoder is employed to achieve the mapping $\mathcal{M}$ from music to text as follows:
\begin{equation}\nonumber
\small
\ell_{m2t} = \sum\nolimits_{i=1}^n ||f(m_i) -\mathcal{M}(g(t_i))||^2.
\end{equation}

\textbf{Group Topology-Preservation Loss (GTP Loss).}
Although the above coordinate model can achieve the music-to-text transformation, we notice that the music representations by the auto-encoder suffer from high non-discrimination, which further prevents the effective transformation. In light of this, we design a novel Group Topology-Preservation (GTP) loss to learn the discriminative music representation. Due to the relatively high discrimination in the text space, we aim to employ the text to guide the music representation generation. 

\begin{figure*}
  \begin{minipage}[c]{0.67\textwidth}
    \vspace{-3mm}
    \includegraphics[width=.99\textwidth]{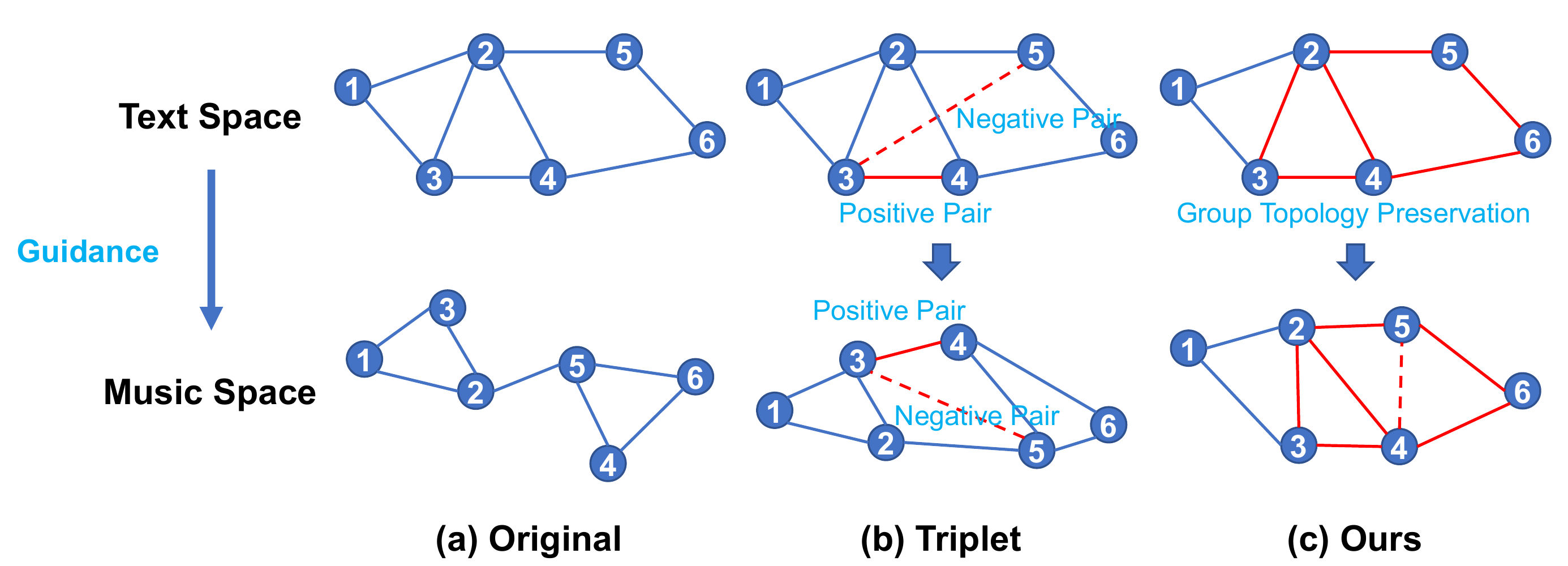}
  \end{minipage}\hfill
  \begin{minipage}[c]{0.32\textwidth}
    \caption{Our group topology-preservation (GTP) loss.
    (a) The structure of data points in the original text and music spaces. 
    (b) The triplet loss is able to enforce the similarity between pairs of points, but fails to maintain the relative positions with other data points. (c) Our GTP loss considers a group reference for each point and aligns the music representations with their structure in text space.}\label{fig:group_regularizer}
  \end{minipage}
 \vspace{-3mm}
\end{figure*}

To better illustrate our point, we provide a visualization of our designed group topology-preservation loss and the comparison with the triplet loss in \Cref{fig:group_regularizer}, where (a) shows the original representation space of music and text derived from the auto-encoders, (b) shows the effect of the triplet loss and (c) shows the effect of our GTP loss. When applying the triplet loss to node \#3 in \Cref{fig:group_regularizer}(b), we observe that nodes \#3 and \#4 get close and nodes \#3 and \#5 are more separated. However, its relative positions as opposed to other data points are still different from the structure in the original text space.
In \Cref{fig:group_regularizer}(c), our proposed GTP loss considers more data points as a group reference and preserves the relative topology among different nodes, thus it can better align the music representations with the structure in text space.

Formally, given a paired tuple $\langle m_i, t_i\rangle$ and a batch $\mathcal{B}=\{\langle m_j, t_j\rangle\}_{j=1}^{k-1}$ containing other $k$$-$$1$ paired tuples as the group reference, our GTP loss can be defined as follows:
\begin{equation}\nonumber
\small
    \ell_{\textup{GTP}} = \sum\nolimits_{i=1}^n || \textup{softmax}(\textup{CON}(\cos(f(m_i),f(m_j)))_{j=1}^{k-1} -\textup{softmax}(\textup{CON}(\textup{BLEU}(t_i,t_j))_{j=1}^{k-1})||^2,
\end{equation}
where $\cos$ is the cosine similarity and BLEU score for the music and text's similarity calculation and $\textup{CON}$ is a concatenation function that flats the input elements and returns a row vector.
Our GTP enforces the given tuple and its group reference share a similar topology structure in both the hidden music space and the original text space, further separating the highly similar music representations. Moreover, the GTP loss not only captures the relationship between the given tuple and other samples in the group reference but also considers the relationship within the group reference. 

\paragraph{Sentiment Alignment Loss.}
To promote the transfer ability of the sentiment from the music domain to the text domain, we use a sentiment classifier to guide the text generation.
Specifically, let $s_i$ be the sentiment label from the ground truth text description $t_i$ and $h(\cdot)$ be a learnable sentiment classifier, 
we then align the sentiment classifier outputs from the original text and the generated text as follows:
\begin{equation}\nonumber
\small
\ell_{\text{s}} = \sum\nolimits_{i=1}^n || s_i - h(g'(f(m_i))) ||^2.
\end{equation}

\paragraph{Overall Objective Function.}
Our overall objective function can be written as follows:
\begin{equation}\nonumber
  \small
    \ell = \ell_{m2m} + \ell_{t2t} + \ell_{m2t} + \alpha \ell_{\textup{GTP}} +  \beta\ell_{\text{s}} + \gamma  \Omega(\Theta),
    \label{eq:overall_loss}  
\end{equation}
where $\Theta$ is the whole learning parameters including $f$, $f'$, $g$, $g'$, $h$ and $\mathcal{M}$, $\Omega$ is the parameter regularizer and $\alpha$, $\beta$, $\gamma$ are trade-off hyperparameters among different losses in the objective function.

\section{Experiments}

\subsection{Experimental Setup}
\label{sec:exp_setup}

We propose five competitive methods and their variants for our task, which are in two main categories: heuristics and pre-trained methods.

\paragraph{Heuristics Methods.} 
In our dataset, each music recording is associated with 4 types of tags, including mode, instrument, tempo and ensemble (in \Cref{tb:data_tag_categories}). 
These tags normally contain information that categorizes the characteristics of the recordings.
Motivated by this observation, we design two simple baselines that are based on heuristics.  

\textsc{Tags ($k$NN)}.  Inspired by the $k$ nearest neighbor clustering, \textsc{Tags ($k$NN)} assigns the text from the closest tags for each piece of music. Specifically, we directly use the text description from a music recording which has the largest number of same tags with our given music. 

\textsc{Tags (representative)}. Alternatively, we first divide all music pieces into separated groups based on four categories of tags. For each tag combination, we select a representative sentence, which has the highest BLEU score within its equivalent classes. These representative sentences can be formed as a text description for given music with the same tag combination.

\paragraph{Pre-trained Methods.}  
We experiment with three representative models for multi-modality translation (discussed in \Cref{sec:related_work}). 
Detailed implementation of these models are in Appendix C.  

\textsc{Encoder Decoder.} Encoder-Decoder model consists of a music CNN encoder and a text transformer decoder. The CNN encoder takes the spectrum of music as input and its output is then directly sent to the text transformer decoder to generate text descriptions. 

\textsc{\textsc{Joint}}. Joint model is based on the assumption that information from different modalities is the same.
A music encoder and a text encoder are used to first encode the music and text inputs. Then the encoded outputs are sent to an adversarial discriminator to identify the encoding source, forcing the alignment of representation from different modalities.
Finally, the encoding vector from both music or text is decoded into text description without modality source information.

\textsc{Coordinate}. Coordinate model representing different modalities in different spaces. To do so, two autoencoders are used for music and text representation. The text decoder is also required to learn a mapping of music to the description text. 
Based on \textsc{Coordinate}, we further propose its variants. 
(1) Contrastive loss~\citep{chen_simple_2020} builds positive samples by pitch shift, time stretch, random mask and noise addition \citep{won_evaluation_2020} to increase representation robustness.
(2) Pairwise loss pulls a pair of music representations together with similar texts.
(3) Triplet loss \citep{schroff_facenet_2015} sets a margin to pull similar positive samples and push other negative samples away.

\textbf{Evaluation}. 
We evaluate the model performance using 5-fold cross-validation.We use the BLEU score \citep{papineni-etal-2002-bleu} as our main evaluation metric to compare our generated captions with the ground truth.

\subsection{Main Results}

We run 3 experimental trails for each split and then report the average and standard deviation of the BLEU scores in \Cref{tb:main_results}.\footnote{We experimented with MusCaps~\citep{9533461} and observed it has very low performance on our dataset.}
Note that the ground text descriptions for classical music are sometimes hard to digest even for humans (see the cases in \Cref{tb:examples}), it is thus not totally surprising that the performance of our model is generally low.

\begin{table*}[t]
\centering
\scriptsize
\caption{Comparisons of different models for our music-to-text generation task using BLEU metric. For all neural-based methods, we run experiments 3 times for each split and report average scores and standard deviations. $k$ in triplet loss refers to different pair numbers used. Our experimental results compared to baseline methods are statistically significant with $p$-value $<$0.001. 
}
\resizebox{0.95\textwidth}{!}{
\begin{tabular}{l|ccccc|c}
\toprule
 & Split \#1 & Split \#2 & Split \#3 & Split \#4 & Split \#5 & \textbf{Average}
 \\\midrule
\textsc{Tags ($k$NN)} & 3.95 & 3.73 & 3.6 & 3.84 & 3.57  & 3.74\\
\textsc{Tags (represent.)} & 3.75 & 3.65 & 4.06 & 3.51 & 3.13 & 3.62\\\midrule
\textsc{Encoder Decoder} & 5.86$_{\pm0.21}$ & 6.31$_{\pm 0.27}$ & 6.36$_{\pm 0.15}$ & 5.77$_{\pm 0.20}$ & 6.59$_{\pm 0.30}$ & 6.18\\
\textsc{Joint} & 5.84$_{\pm 0.66}$ & 6.31$_{\pm0.24}$ & 6.48$_{\pm0.53}$ & 5.83$_{\pm0.27}$ & 6.27$_{\pm0.27}$ & 6.15\\
\textsc{Coordinate} & 6.41$_{\pm0.34}$ & 6.70$_{\pm0.17}$ & 6.20$_{\pm0.21}$ & 6.10$_{\pm0.11}$ & 6.66$_{\pm0.24}$ & 6.42 \\\midrule
\textsc{Coor.} + Contrastive loss & 5.71$_{\pm0.16}$ & 4.80$_{\pm0.18}$ & 6.74$_{\pm0.24}$ & 6.11$_{\pm0.11}$ & 6.98$_{\pm0.47}$ & 6.07\\
\textsc{Coor.} + Pairwise loss & 6.20$_{\pm0.32}$ & 6.76$_{\pm0.10}$ & 6.53$_{\pm0.29}$ & 5.97$_{\pm0.20}$ & 6.53$_{\pm0.21}$ & 6.40 \\
\textsc{Coor.} + Triplet loss ($k$=1) & 6.60$_{\pm0.15}$ & 6.47$_{\pm0.13}$ & 6.74$_{\pm0.07}$ & 6.07$_{\pm0.13}$ & 6.95$_{\pm0.14}$ & 6.56 \\
\textsc{Coor.} + Triplet loss ($k$=32) & 6.36$_{\pm0.28}$  & 6.59$_{\pm0.07}$  & 6.50$_{\pm0.20}$ & 5.94$_{\pm0.24}$ & 6.71$_{\pm0.21}$ & 6.42 \\
\textsc{Coor.} + Sentiment loss &  6.49$_{\pm0.16}$  & 6.93$_{\pm0.12}$  & 6.59$_{\pm0.38}$ & 6.01$_{\pm0.27}$ & 6.88$_{\pm0.25}$ & 6.58\\
\midrule
\textsc{Ours} w/o Sentiment loss & 6.94$_{\pm0.19}$ & 6.71$_{\pm0.15}$ & 6.91$_{\pm0.25}$ & 6.37$_{\pm0.24}$ & 6.87$_{\pm0.16}$ & 6.76 \\
\textsc{Ours}  &  7.04$_{\pm0.09}$ & 6.79$_{\pm0.19}$ & 6.66$_{\pm0.36}$ & 6.55$_{\pm0.19}$ & 7.05$_{\pm0.19}$ & \textbf{6.82}\\
\bottomrule
\end{tabular}
}
\label{tb:main_results}
\end{table*}

We observe that heuristics baseline methods have lower performance compared to learning-based models. 
It is intuitive as heuristics-based methods are not able to generate text descriptions that are adjusted to given music contents. 
Among all three representative models for multi-modal tasks, \textsc{Coordinate} achieves the best performance with an average BLEU score of 6.42. 
When further adding different losses to the training objective of the coordinate model, we observe similar performance for using triplet loss and even drops for contrastive loss and pairwise loss; while there is a significant performance improvement when using our proposed GTP loss, achieving the highest BLEU score of 6.82. 
Some music recordings on earsense have a second comment by another commentator. 
We collect these data as another dataset with 50 records to validate our models. The BLEU score of the baseline \textsc{Coordinate} is 8.54 and our GTP loss model is 8.81.

\begin{figure*}[t]
\centering
\subfigure[Pretrained music autoencoder]{
\centering
\includegraphics[width=0.3\linewidth]{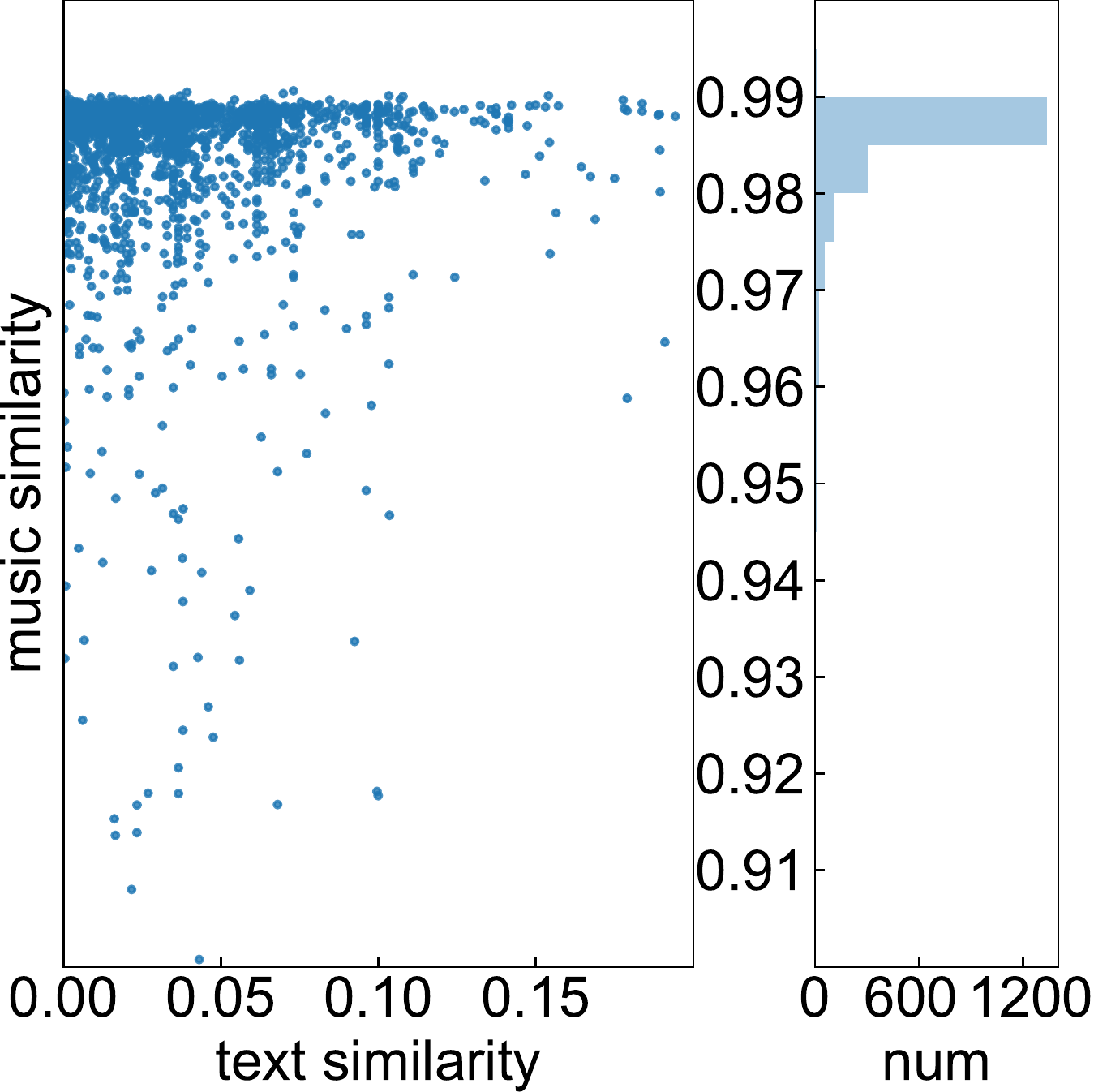}\vspace{1mm}
}%
\subfigure[\textsc{Coor.} + Pairwise loss]{
\centering
\includegraphics[width=0.3\linewidth]{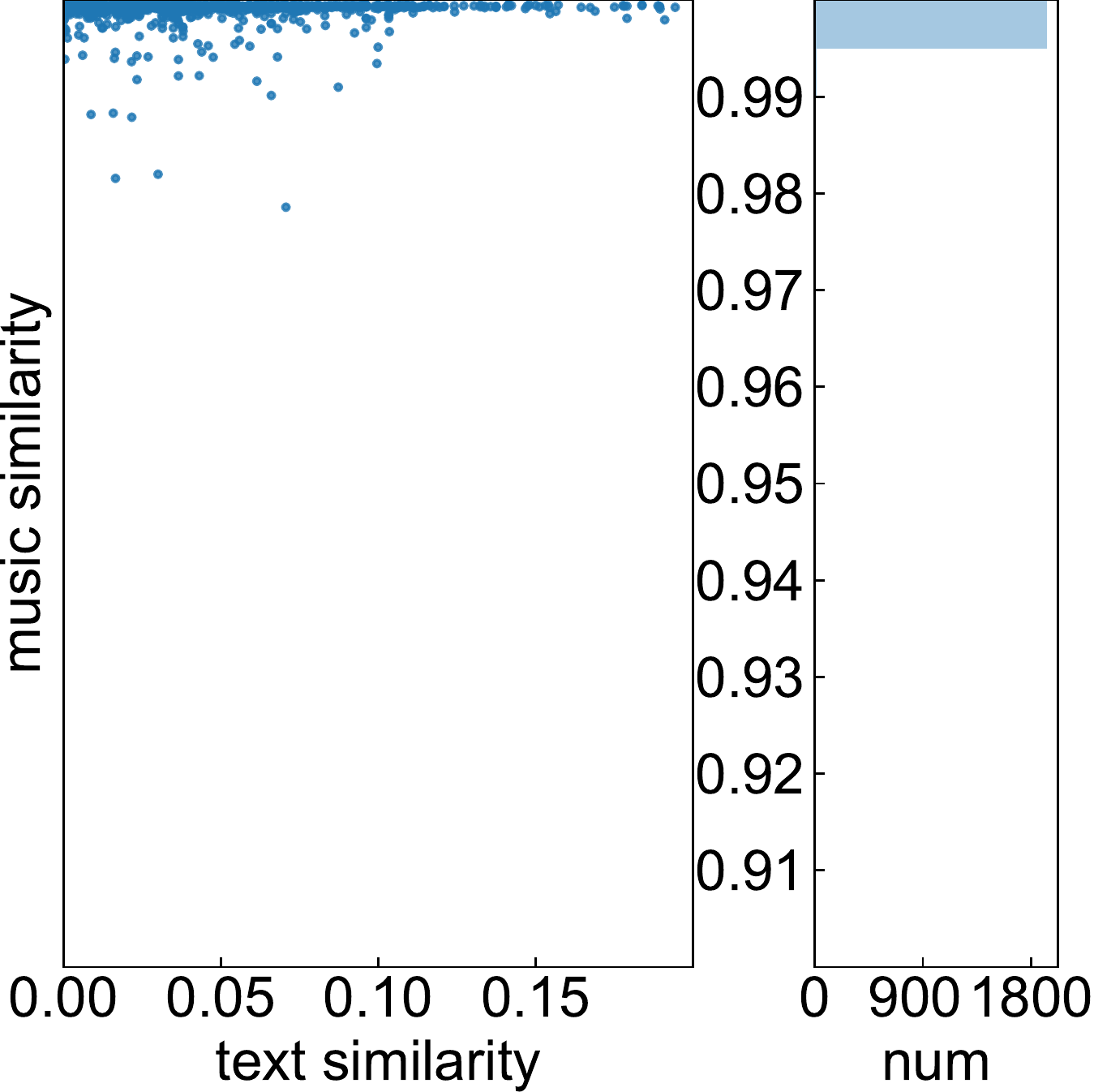}\vspace{1mm}
}
\subfigure[\textsc{Ours}]{
\centering
\includegraphics[width=0.3\linewidth]{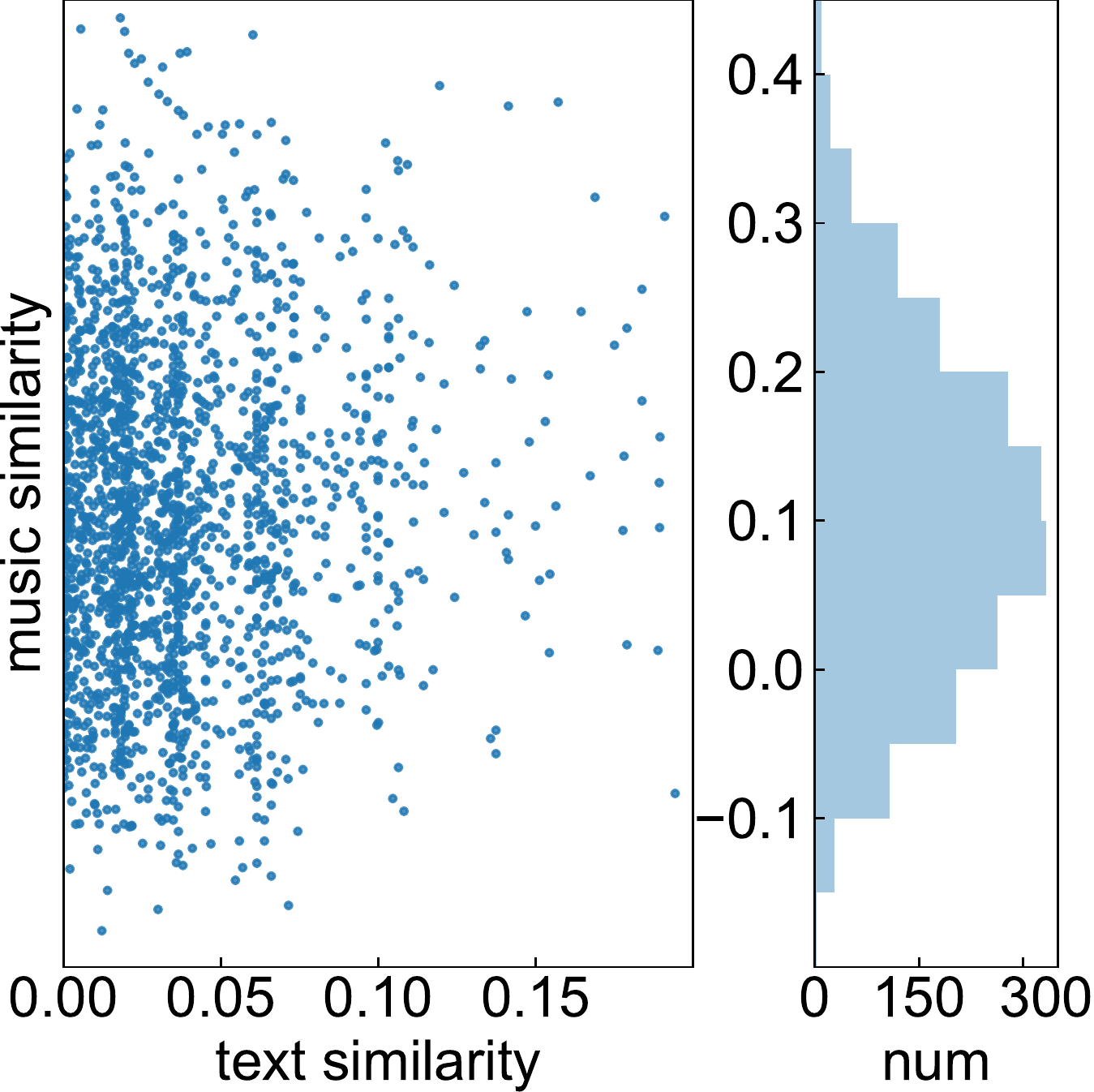}\vspace{-8mm}
}
\vspace{-2mm}
\caption{Music and text similarities of pairs of music-text tuples and histograms on the music similarity.
(a) Pretrained music autoencoder; (b) \textsc{Coordinate} with pairwise loss (triplet loss has a similar trend); (c) \textsc{Ours}.} 
\label{fig:text_music_scatter}
\end{figure*}

To better interpret this experimental result, in \Cref{fig:text_music_scatter} we plot the music and text similarities of a pair of different music-text tuples and the histograms on the music similarity when applying different losses.
The music similarity is calculated by the cosine similarity on the learned representation and the text similarity is calculated by the BLEU score on the original texts.
The output representations from pre-trained music autoencoder have high similarity, with the majority pairs above 0.90 in \Cref{fig:text_music_scatter}(a).
It is because the high frequency of all music is almost blank while classical music also has a similar low frequency with several limited number of instruments.
When using pair loss, the music representations simply gather together and become indistinguishable (in \Cref{fig:text_music_scatter}(b)).
Meanwhile, using our proposed GTP loss helps maintain a structure among music representations.
As shown in \Cref{fig:text_music_scatter}(c), the music similarities vary from \mbox{-0.05} to 0.4 and the range is much wider than that of other models.
Also, the representations spread out and become more distinguishable, which naturally helps the performance of text generation.
It validates the motivation of our GTP loss that when considering more data points as a group reference, our model captures the relative topology among different nodes, thus can better accomplish the transformation from different modalities.

\begin{table}[t]
    \centering
    \footnotesize
    \caption{
    Generated text descriptions between different models given a classical music piece, followed by their sentiments [joy/neurtal/sadness] in the end. We use different colors to highlight phrases: blue for noun phrases that is objective facts and yellow for subjective adjective phrases. 
    }\vspace{1mm}
    \resizebox{\textwidth}{!}{
    \begin{tabular}{l|p{0.99\textwidth}}
    \toprule        
        \multirow{8}{*}{\makecell[c]{Case \#1}}
        & \textbf{Reference}: the \hlyellow{rousing} \hlblue{allegro} vivace finale begins with a \hlyellow{sweetly} singing adagio introduction that very briefly seems to recall the extended introductions of the earlier op. 5 sonatas and provides the only real hint of a slow movement that will otherwise have to wait for the last \hlblue{cello} \hlblue{sonata} for its fully independent flowering. [joy]\\
        & \textbf{\textsc{Coordinate}}: the movement, \hlblue{allegro} moderato, begins with a fugue with a fugue. [neutral]\\
        & \textbf{\textsc{Ours}}: the movement, \hlblue{allegro} moderato, begins with a theme of the main theme of the main theme in the \hlblue{cello}. the main theme is a kind of the middle section, the work of the movement. the music is a fugue with a bit more \hlyellow{lyrical} second theme to the \hlblue{sonata} form of pizzicato. [joy]\\\midrule
        \multirow{7}{*}{\makecell[c]{Case \#2}} & 
        \textbf{Reference}: the movement, \hlblue{allegro molto} e con brio, is based on two subjects, \hlyellow{the first is bright and lively} while \hlyellow{the second is dreamy} with an improvisational aura. the development is ingenious and an exciting coda caps off this first rate work. [joy]\\
       & \textbf{\textsc{Coordinate}}: the movement, \hlblue{allegro moderato}, begins with a theme in the main theme. [joy]\\
        & \textbf{\textsc{Ours}}: the movement, \hlblue{allegro molto}, is a series of variations that it \hlyellow{begins with a march - like it is a scherzo} and a sense of the main theme, in the movement. it is a very \hlyellow{romantic second theme}, a sense of the movement. [joy] \\\midrule
        \multirow{5}{*}{\makecell[c]{Case \#3}}
        & \textbf{Reference}: the movement, \hlblue{allegro moderato}, begins with a \hlyellow{somewhat passionate, very characteristic} quartet - like melody which leads to an ingratiating, \hlyellow{lyrical} second theme. [joy] \\
        & \textbf{\textsc{Coordinate}}: {the movement, \hlblue{allegro}, begins with a lengthy introduction.} [neutral]\\  
        & \textbf{\textsc{Ours}}: { the movement, \hlblue{allegro}, begins with a \hlyellow{lively allegro con brimming} with a long series of the main theme. it begins with a very \hlyellow{romantic} melody, the music of the movement.} [joy] \\
    \bottomrule
    \end{tabular}
    }
    \label{tb:examples}
\end{table}

\paragraph{Case Study.}
We present several examples in \Cref{tb:examples}, showing the comparisons between ground truth descriptions and the generated ones from \textsc{Coordinate} with or without our proposed GTP loss. 
We observe \textsc{Coordinate} and \textsc{Ours} are able to identify basic characteristics of the input musics, while our model in general could capture more subtle differences.
For example, in Case \#1, both models correctly identify the tempo ``\textit{allegro},'' while our model further captures the music instrument of ``\textit{cello}'' and genre of ``\textit{sonate.}''
In Case \#2, our model captures two themes in one movement, where the ``\textit{bright and lively}'' first theme is recognized as a ``\textit{march like a scherzo,}'' and the ``\textit{dreamy}'' second theme is described to be ``\textit{romantic}.''

It can be observed that, for objective facts, models can simply and correctly tell the correct term, while our model can further capture more details. 
When it comes to subjective description, our model could capture the sentiment orientation of music themes (see \Cref{sec:analysis} for further analyses of the sentiment transfer ability between different modalities), although it is still a challenging task due to expression diversity.
We also notice some grammar mistakes and fixed patterns appear in generated sentences. It seems that even with sufficient pre-training, it is still hard to learn a professional expression of both correctness and fluency from our collected dataset with a limited size.

\subsection{In-depth Analyses}
\label{sec:analysis}

\begin{figure*}[t]
\centering
\subfigure[Number of samples $k$ in GTP]{
\centering
\includegraphics[width=0.298\linewidth]{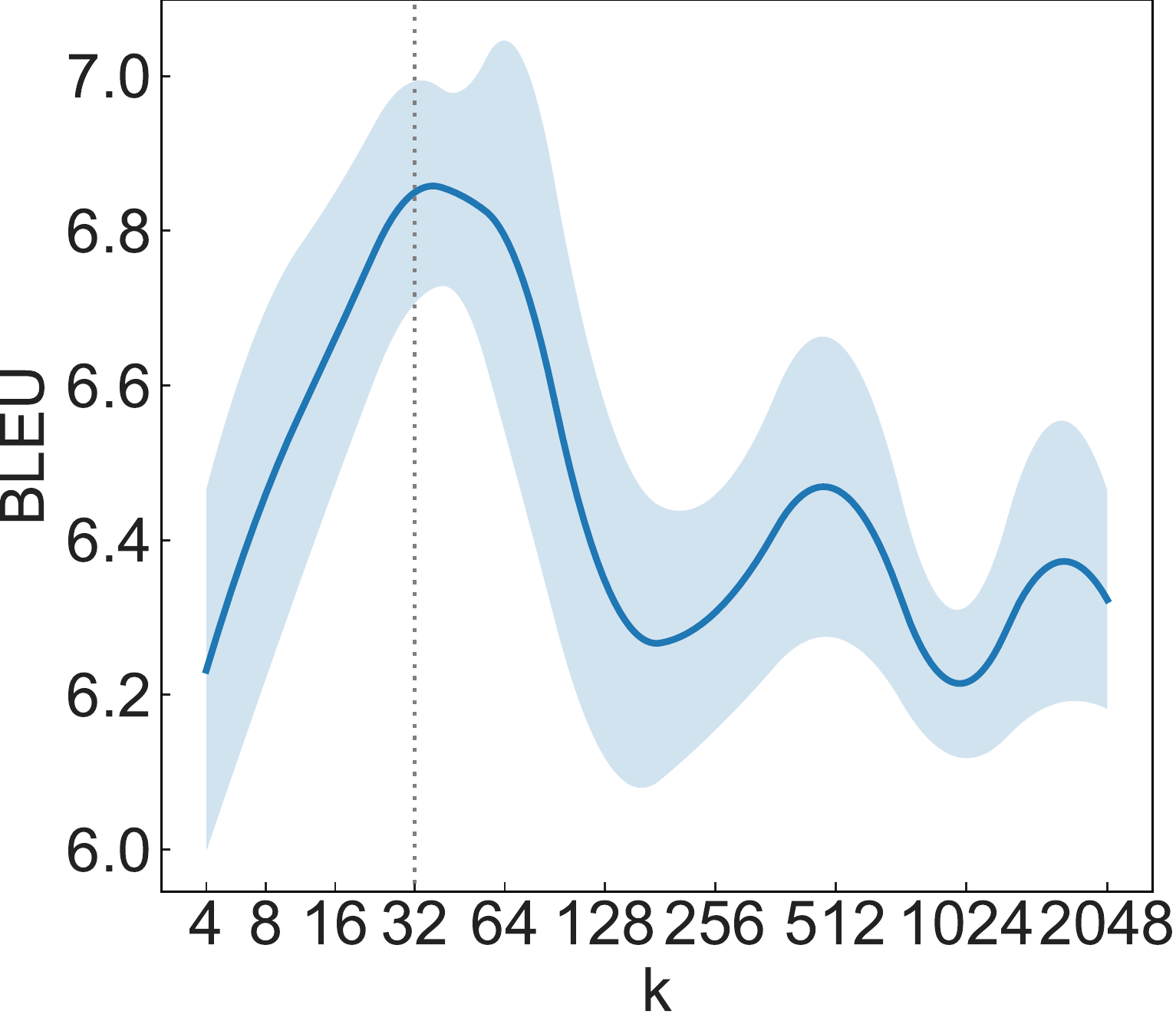}\label{fig:k}
}
\subfigure[Parameter sensitivity]{
\centering
\includegraphics[width=0.298\linewidth]{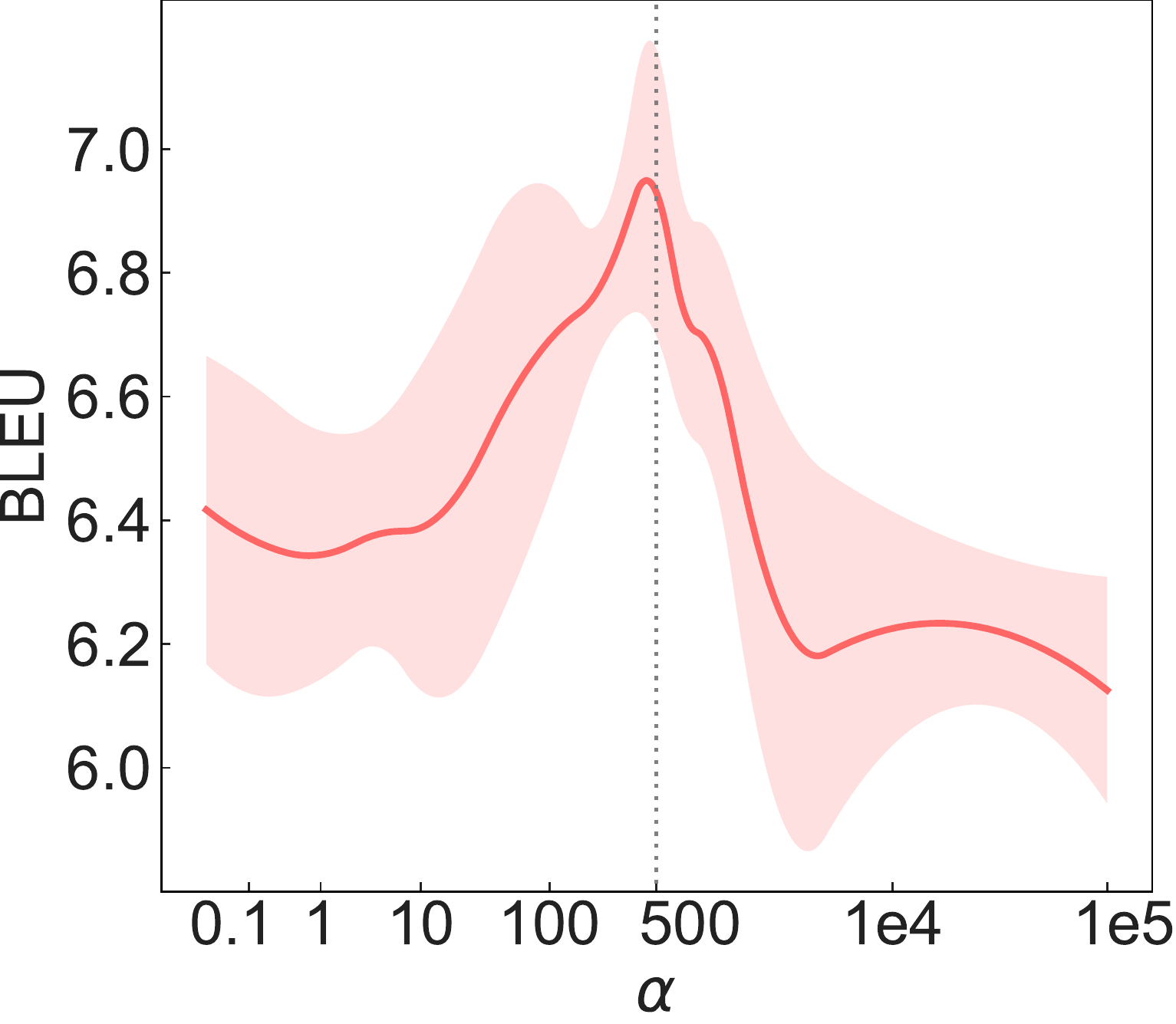}\label{fig:alpha}
}\hspace{-1.5mm}
\subfigure[Sentiment distribution]{
\raisebox{2mm}{
\centering
\includegraphics[width=0.344\linewidth]{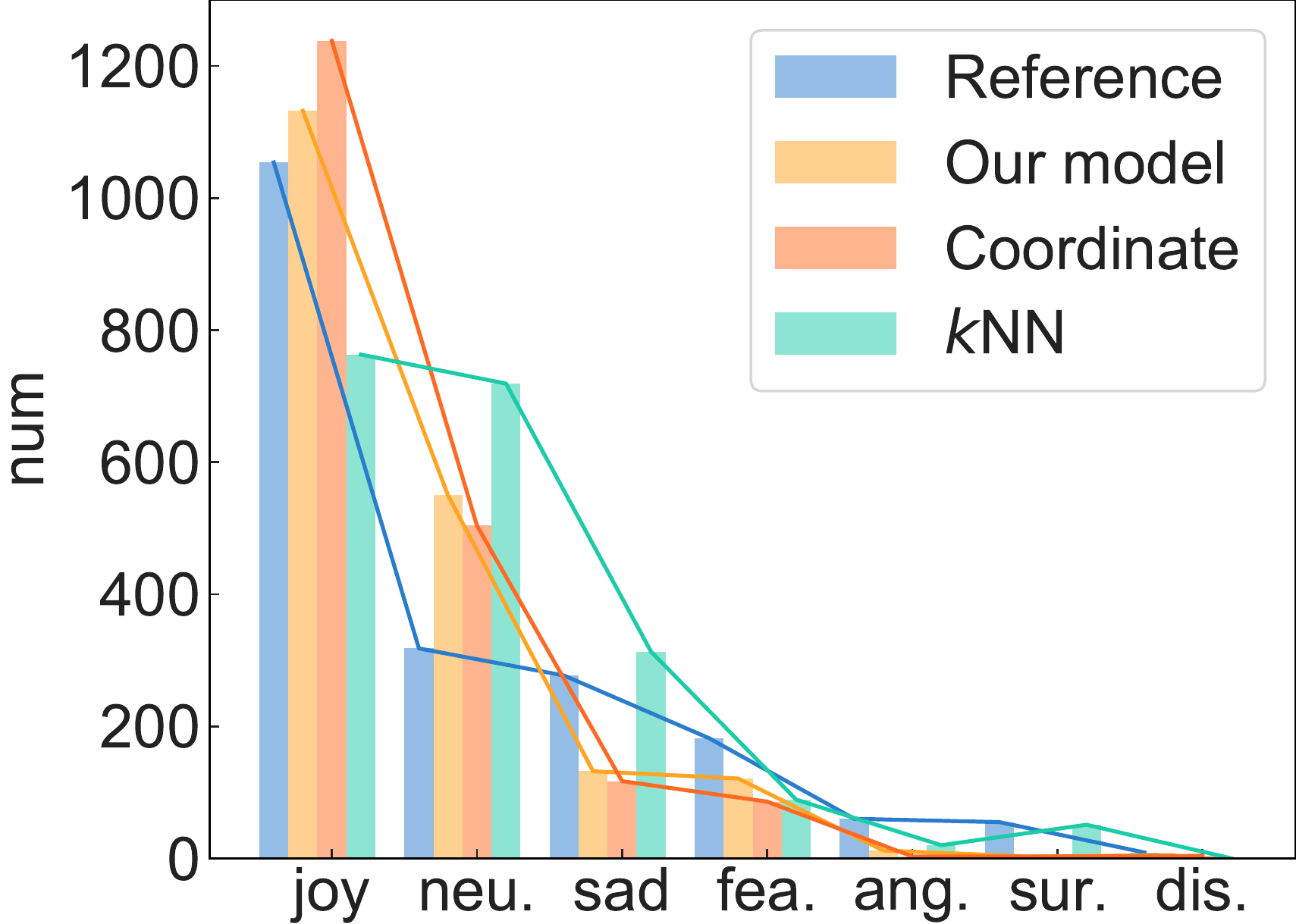}}
}
\vspace{-2mm}
\caption{(a) BLEU score when varying the number of neighbors $k$. (b) Parameter sensitivity on $\alpha$. (c) Sentiment distribution of our generated texts and the original reference texts.} 
\label{fig:analysis}
\end{figure*}

\textbf{Effects of Hyperparameters.}
We are interested in the effect of the number $k$ of samples considered in our GTP loss. 
To do this, we vary the number of $k$ from 4 to 2,048 and plot the BLEU scores in \Cref{fig:k}.
We observe that the performance of our model first increases and achieves its peak when $k$ is chosen to be 32; the performance then drops when $k$ is larger.
It provides empirical evidence that considering more information from other data points do help improve the model performance. However, a larger $k$ is not always better for which will make older samples in the momentum encoder queue more outdated.
We observe a similar pattern when varying the scale of the hyperparameter $\alpha$ from 1e-2 to 1e5 in \Cref{fig:alpha}. The model yields the best performance when $\alpha$ = 500.

\paragraph{Transfer Ability of Sentiment.}
Since our task is music-to-text synaesthesia, we care about whether our model could learn a transfer capability from the music domain to the text domain. To explore this, we analyze whether the generated text descriptions are able to reflect the sentiment from the original ground truth text. 
We select a popular sentiment classifier\footnote{ \scriptsize{\url{https://huggingface.co/j-hartmann/emotion-english-distilroberta-base}}} that could detect a variety of 7 sentiment categories and then apply this classifier to identify the sentiment in reference text and text descriptions generated by our model. \Cref{fig:analysis}(c) shows the sentiment distribution of our generated texts and the original reference texts. We observe a skewed distribution that a majority of descriptions from both references and generated models are classified as ``joy,'' while the left others are tagged as having sentiments like ``neutral,'' ``sadness,'' and ``fear.'' These experimental results are consistent with our intuitions that classical music normally brings people with happiness and joy. We observe that the distributions for the sentiment labels from the reference and our model are similar, with a Pearson correlation coefficient of up to 0.96.

\section{Conclusion}

In this paper, we proposed a novel task of music-to-text synaesthesia, the goal of which is to generate text descriptions for a given music piece.
As current existing datasets normally do not contain semantic descriptions for music pieces, we collected a new dataset that contains 1,955 classical music recordings with professional text descriptions.
We also proposed a group topology-preservation loss that preserves relative group topology among music and text, given the fact that classical music is non-discriminative.
We conducted experiments on five heuristics or pre-trained competitive baseline methods, based on existing multi-modal literature.
Compared with other sample-level constraint losses, our model with GTP loss extracted better music representation and generated better text descriptions.
We further performed extensive experiments including case studies and sentiment analysis to demonstrate the effectiveness of our proposed music-to-text synaesthesia model.

\section*{Limitations}

One limitation of our collected dataset is that its size is relatively small. 
It also only covers western classical music.
The model trained on our dataset thus may not generalize well to other types of music.  
However, to the best of our knowledge, it has been the largest dataset containing high-quality music-text description pairs that we could collect.

\newpage

\bibliography{ref}
\bibliographystyle{unsrtnat}

\newpage
\appendix

\setcounter{table}{0}
\renewcommand{\thetable}{A\arabic{table}}
\setcounter{figure}{0}
\renewcommand{\thefigure}{A\arabic{figure}}

\section{Data Analysis}
\label{sec:data_stats}

\subsection{Dataset Statistics}

\begin{figure*}[h!]
\centering
\subfigure[Music Length]{
\centering
\includegraphics[width=.29\textwidth]{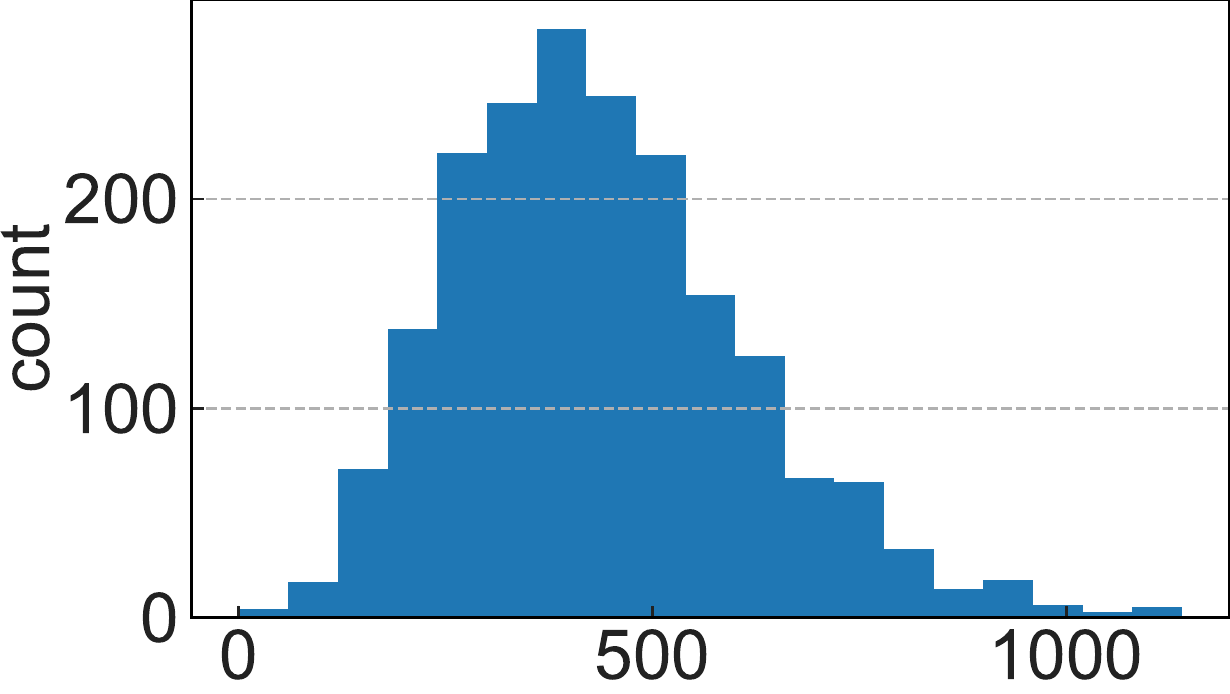}
}
\subfigure[Token Numbers]{
\centering
\includegraphics[width=.29\textwidth]{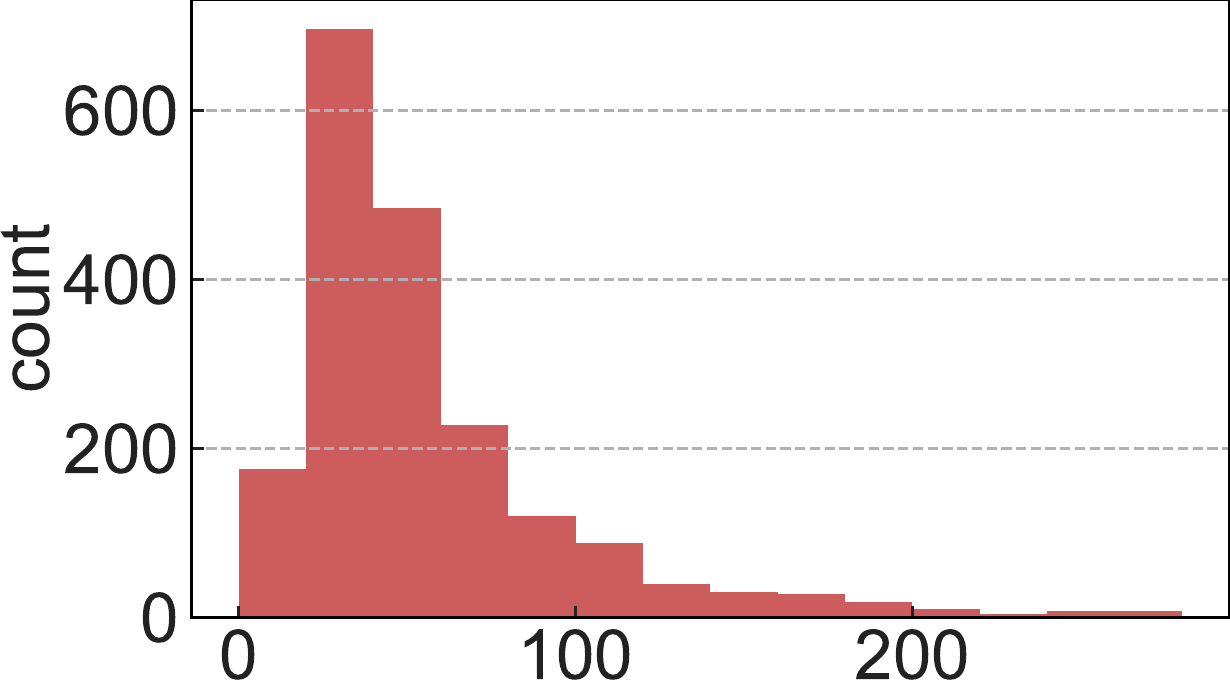}
}
\vspace{-2mm}
\caption{Distribution of music length and token nums}
\label{fig:data_length_distribution}
\end{figure*}

\begin{figure}[h!]
    \centering
    \includegraphics[width=.88\textwidth]{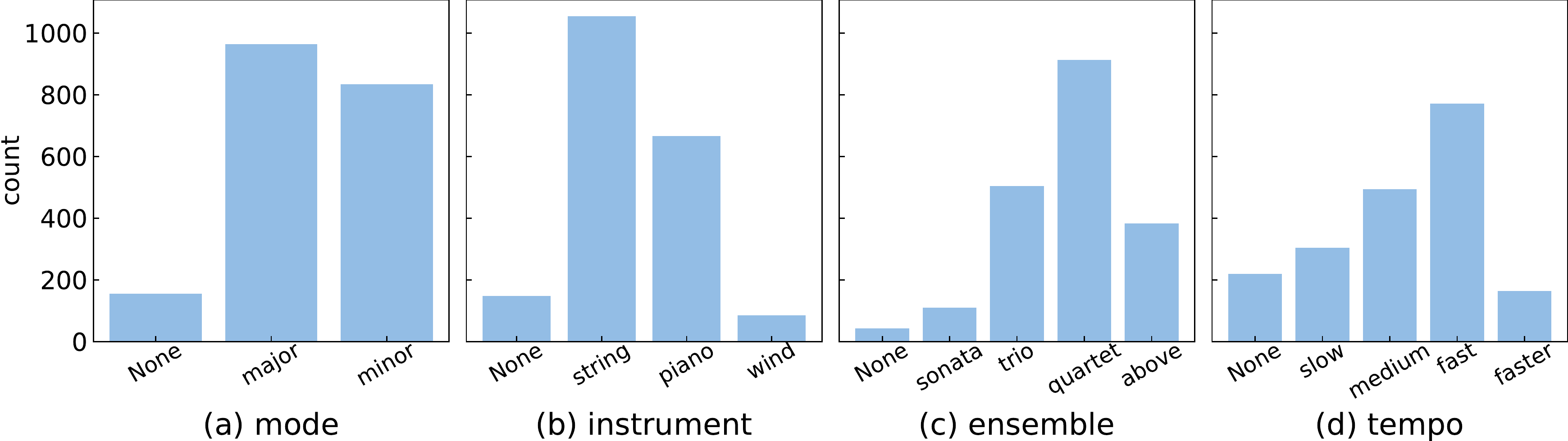}\vspace{-2mm}
    \caption{Distribution of four tag categories}
    \label{fig:tag_distribution}
\end{figure}

We summarize the statistics of our newly collected dataset in \Cref{tb:data_stats,tb:data_tag_categories}, \Cref{fig:data_length_distribution,fig:tag_distribution}. There are several advantages and unique features for this dataset: (1) Music pieces are semantically hard enough, which are provide space for generating text with rich information. (2) The descriptive texts are generated by music professionals, thus are freestyle and diverse.

\begin{table}[h!]
\resizebox{0.95\textwidth}{!}{
\centering
\begin{minipage}{.5\linewidth}
\centering
\small
\caption{Statistics of our collected dataset}\label{tb:data_stats}
\begin{tabular}{lc}
\toprule
\# of unique music compositions & 786\\
\# of unique composers & 304 \\
\# of music-text pairs & 2,380\\
\midrule
\# avg. of tokens per text description & 45.7 $\pm$ 41.2 \\
\# avg. of each music piece (seconds) & 305 \\
\bottomrule
\end{tabular}
\end{minipage}
\quad
\begin{minipage}{.5\linewidth}
\centering
\small
\caption{Four categories of tags in our dataset}\label{tb:data_tag_categories}
\vspace{-.05cm}
\begin{tabular}{ll}
\toprule
Tag &  Type\\
\midrule
mode & major, minor\\
instrument & string, wind, piano \\
tempo & slow, medium, fast, super faster\\
ensemble & sonate, trio, quartet, quintet or more\\
\bottomrule
\end{tabular}
\end{minipage}
}
\end{table}

\begin{figure}[ht]
    \centering
    \includegraphics[width=0.35\linewidth]{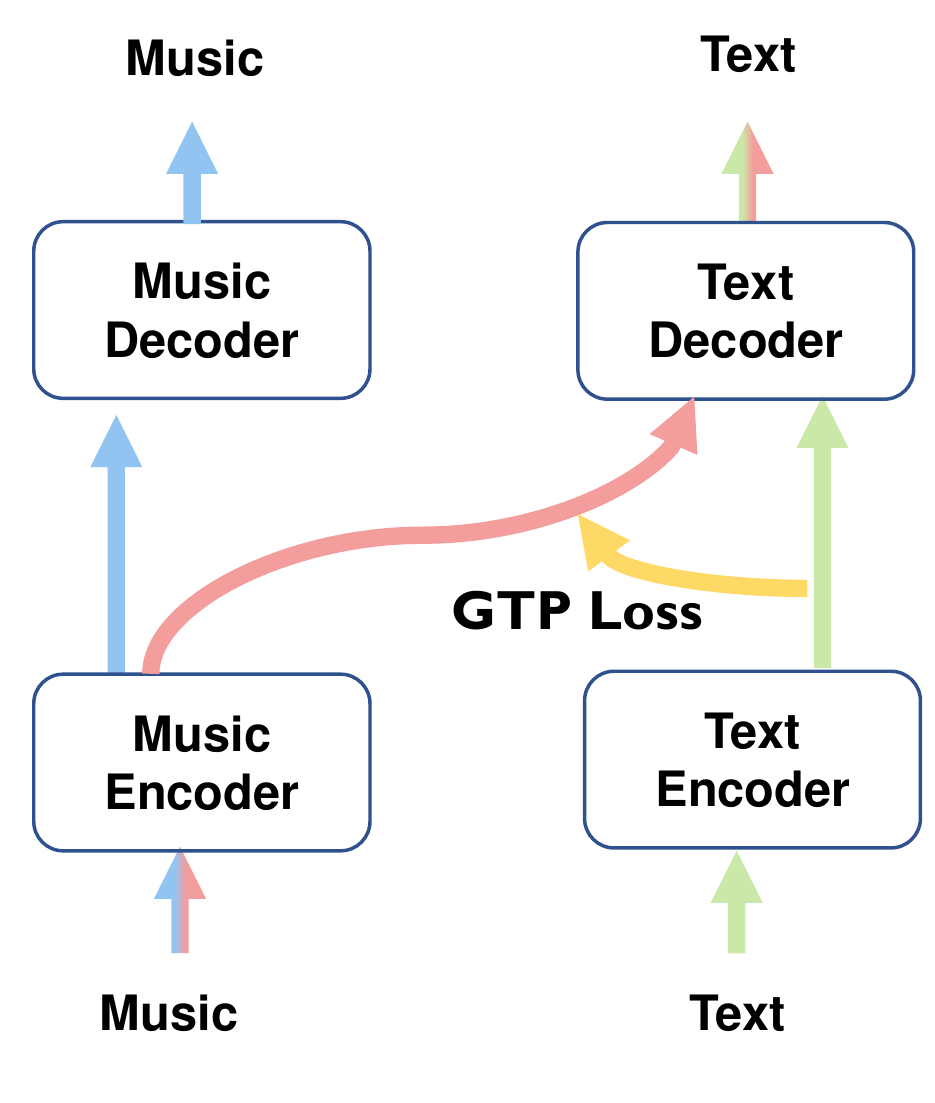}
    \vspace{-2mm}
    \caption{Framework for our music-to-text synaesthesia task. Different colors denote forward data streams in one training step.} 
    \label{fig:framework}
\end{figure}

\section{Model Architectures}

\begin{figure*}[!ht]
\centering
\subfigure[\textsc{Enco. Deco.}]{
\centering
\includegraphics[height=5.5cm]{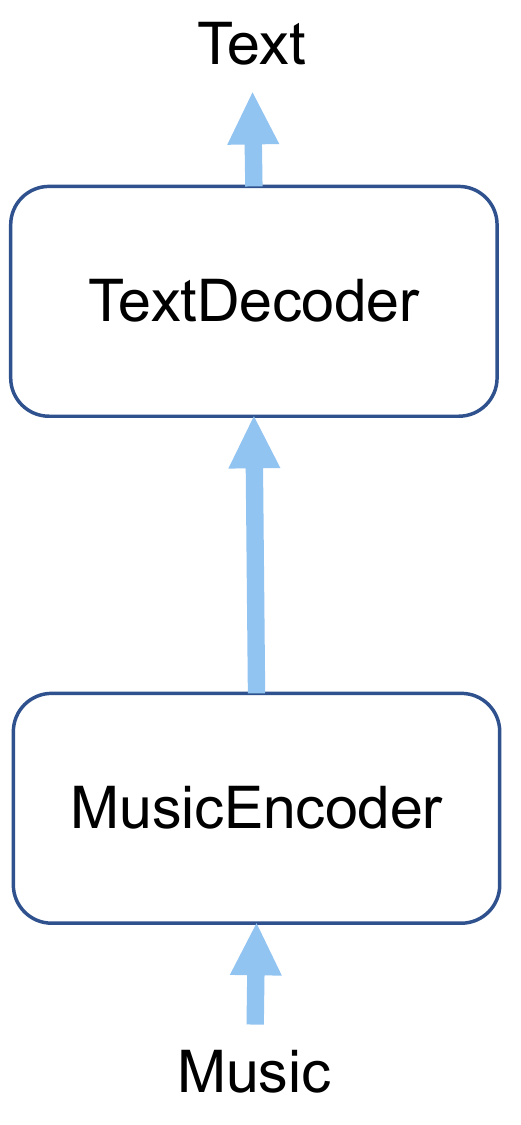}
}\hspace{4mm}
\subfigure[\textsc{Joint}]{
\centering
\includegraphics[height=5.5cm]{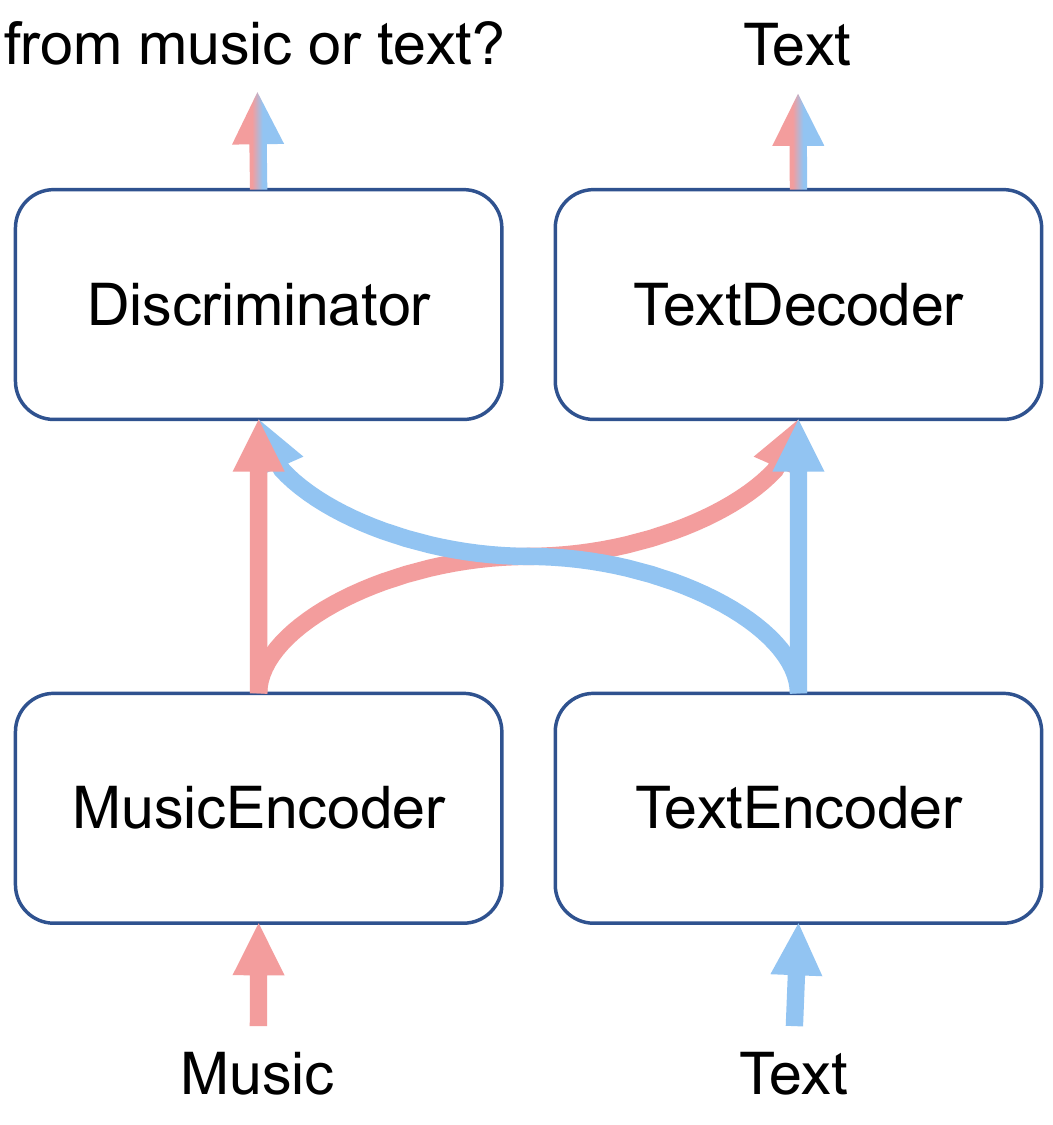}
}\hspace{4mm}
\subfigure[\textsc{Coordinate}]{
\centering
\includegraphics[height=5.5cm]{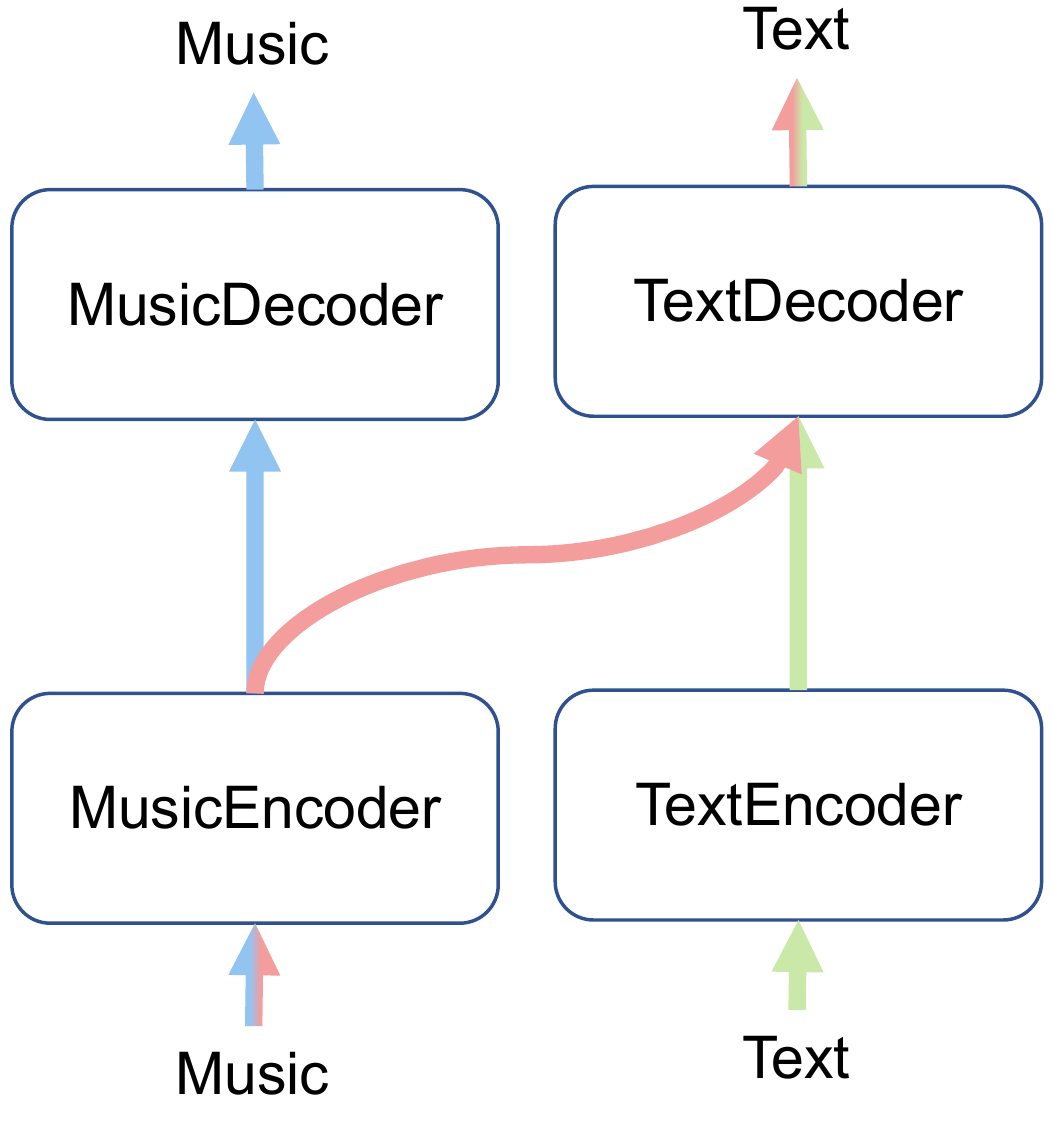}
}
\vspace{-4mm}
\caption{Structures of three commonly used framework for multi-modality translation (details in \Cref{sec:exp_setup}).}
\label{fig:structures}
\end{figure*}

We present our proposed framework for music-to-text synaesthesia task in \Cref{fig:framework}.
Our proposed model is based on the coordinate model, which achieves the best performance among all three commonly used frameworks for multi-modal translation in our pilot studies. 
We visualize the structure of all baseline models in \Cref{fig:structures}.

\section{Implementation}

From the music side, the original classical music is first converted into a spectrogram which has a frequency axis of 256 dimensions and a flexible temporal axis.
It is then encoded into $T\times768$ by CNN encoder, where $T$ is the music length in seconds. Both the music encoder $f(\cdot)$ and decoder $f'(\cdot)$ contain 8 layers. From the text side, the text descriptions are tokenized by BERT tokenizer and encoded into $L\times768$, where $L$ is the number of tokens. The text encoder $g(\cdot)$ and decoder $g'(\cdot)$ consist of 5 layers. As the input music and text are of varying length, both music and text encodings are divided into 20 segments evenly, and mean pooled into $20\times768$ vectors as a fixed-length representation. The same number of layers for music and text encoders are used for \textsc{Encoder Decoder} and \textsc{Joint} baselines.

\begin{wrapfigure}[17]{r}{0.37\textwidth}
\centering
\vspace{-4mm}
\includegraphics[width=.37\textwidth]{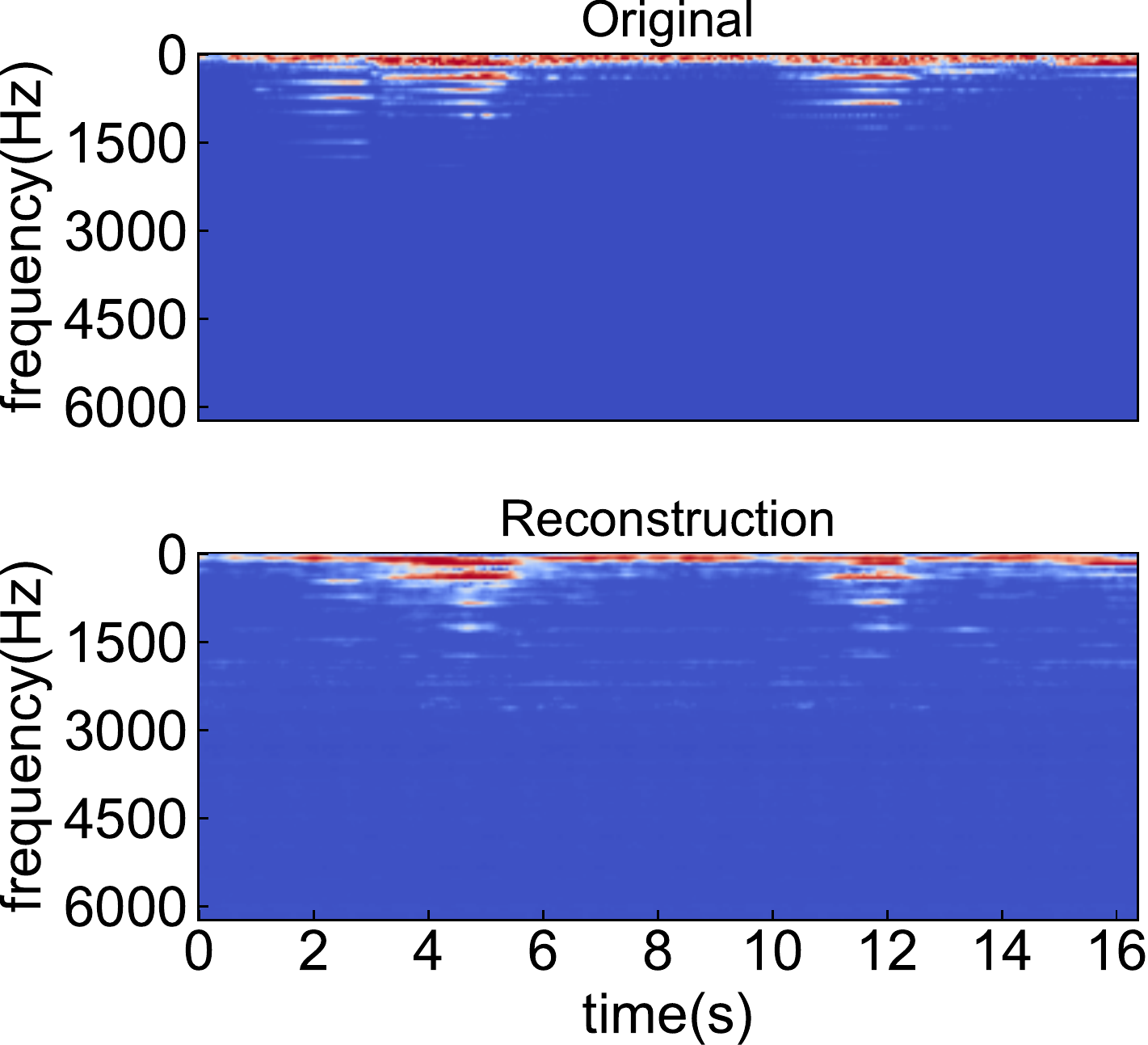}
\vspace{-6mm}
\caption{Spectrum diagrams from the original music (top) and reconstruction (bottom) of our pre-trained music autoencoder.} 
\vspace{-4mm}
\label{fig:wave_recon}
\end{wrapfigure}

For our model, we first pre-train music and text auto-encoders on our constructed dataset.
To prevent data leakage, the text auto-encoder is pre-trained without the test data. Moreover, these music and text auto-encoders capture a good representation in the reconstruction task. \Cref{fig:wave_recon} shows the spectrum diagrams of the original and reconstruction music from music representation, which contains the main information of original music. Meanwhile, the text-autoencoder reconstructs from representation with an average BLEU score of 79.66. 
During training, we set the learning rate to be 5e-5 and the batch size is 8, implemented by gradient accumulation. The hyperparameters in our model are set to be $\alpha$ = 500, $\beta$ = 5 and $k$ = 32. 
For our sentiment loss, we set $\gamma$ = 0.25. 
We choose a popular open-source sentiment model as our sentiment classifier.\footnote{ \scriptsize{\url{https://huggingface.co/j-hartmann/emotion-english-distilroberta-base}}} 
Considering the computational budget, our $k$ grouped sample representations are implemented by a queue according to MOCO \citep{he_momentum_2020}, where we employ a momentum encoder with $momentum$ = 0.999.

\end{document}